\documentclass[11pt]{article}
\usepackage{amsmath}
\usepackage{amssymb}
\usepackage{graphicx}

\def \aa{A\&A }

\def \jgr{J. Geophys. Res. }

\begin{document}

\setcounter{figure}{0}
\setcounter{table}{0}
\setcounter{footnote}{0}
\setcounter{equation}{0}

\vspace*{0.8cm}

\noindent {\Large ANALYSIS AND COMPARISON OF PRECISE LONG-TERM NUTATION SERIES, STRICTLY DETERMINED WITH OCCAM 6.1 VLBI SOFTWARE}

\vspace*{1cm}

\noindent\hspace*{1.5cm} G. BOURDA$^{1~2~3}$, J. BOEHM$^3$, R. HEINKELMANN$^3$, H. SCHUH$^3$\\
\noindent\hspace*{1.5cm} $^1$ L3AB-UMR5804, Observatoire de Bordeaux, FRANCE\\
\noindent\hspace*{1.5cm} $^2$ SYRTE-UMR8630, Observatoire de Paris, FRANCE\\
\noindent\hspace*{1.5cm} $^3$ Institute of Geodesy and Geophysics, Vienna University of Technology, AUSTRIA\\
\noindent\hspace*{1.5cm} e-mail: Geraldine.Bourda@obspm.fr\\

\vspace*{0.8cm}

\noindent {\large 1. INTRODUCTION}

\smallskip

The IAU/IUGG Working Group on "Nutation for a non-rigid Earth", led by V\'{e}ronique Dehant, won the European Descartes Prize in 2003, for its work developing a new model for the precession and the nutations of the Earth. This model (MHB2000, Mathews et al. 2002) was adopted by the IAU (International Astronomical Union) during the General Assembly in Manchester, in 2000. It is based (i) on some improvements for the precession model (with respect to the previous one of Lieske et al. 1977) owing to the VLBI technique (Very Long Baseline Interferometry), and (ii) on a very accurate nutation model, close to the observations. With this prize, the Descartes nutation project could offer for international scientists some grants, to be used for further improvements of the precession-nutation Earth model. At the IGG (Institute of Geodesy and Geophysics), with the OCCAM 6.1 VLBI analysis software and the best data and models available, we re-analyzed the whole VLBI sessions available (from 1985 till 2005) solving for the Earth Orientation Parameters (EOP). In this paper we present the results obtained for the EOP and more particularly for the nutation series. We compare them with the other IVS (International VLBI Service) analysis centers results, as well as with the IVS combined EOP series from the analysis coordinator. The series are in good agreement, except for the polar motion coordinates that show a shift with respect to the other ones and that we discuss here. Finally, we analyse the nutation series in the framework of the free core nutation (FCN) effect.

\vspace*{0.8cm}

\noindent {\large 2. COMPUTATION}

\smallskip

In this study we use 2944 VLBI sessions from 1982 till 2004, described as suitable for determining the EOP by the IVS Analysis Coordinator. They are computed at the IGG, from the NGS-format files corrected on the basis of the ECMWF meteorological data, in order to improve the height component of the stations and the determination of the tropospheric parameters. Clock breaks and reference clocks were investigated and taken into account for each session (Heinkelmann et al., 2005). The VLBI analysis software used at the IGG is OCCAM 6.1, in which the classical least-squares method based on the Gauss-Markov model (Koch, 1997) is implemented. We used (i) the terrestrial reference frame ITRF2000, (ii) the Vienna Mapping Function (VMF; Boehm \& Schuh 2004), and (iii) a cut-off elevation angle for the troposphere set to 5 degrees. We estimated atmospheric gradients, clock parameters and zenith delays, as the same time we solve for the 5 EOP: one per session, and its rate for the pole and UT1. But we do not estimate the stations and sources position. \\


\noindent {\large 3. DISCUSSION}

\smallskip

\begin{figure}[h]
\begin{minipage}[c]{.48\linewidth}
\begin{center}
\includegraphics[scale=0.38]{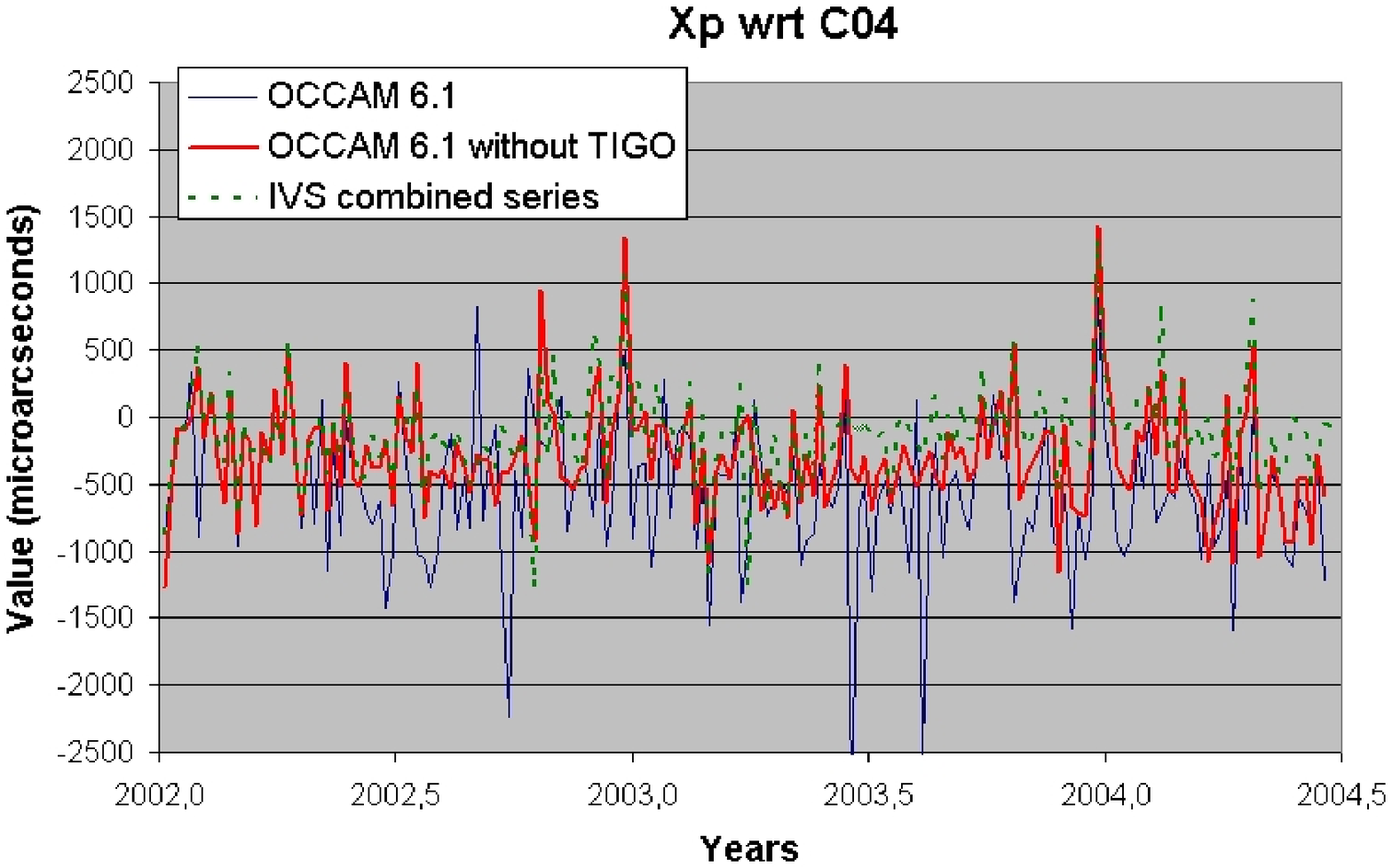}
\caption{Polar motion $x_p$ component, obtained with OCCAM 6.1 and compared to the IVS combined series.}\label{fig:Xp}
\end{center}
\end{minipage} 
\hfill
\begin{minipage}[c]{.48\linewidth}
\begin{center}
\includegraphics[scale=0.38]{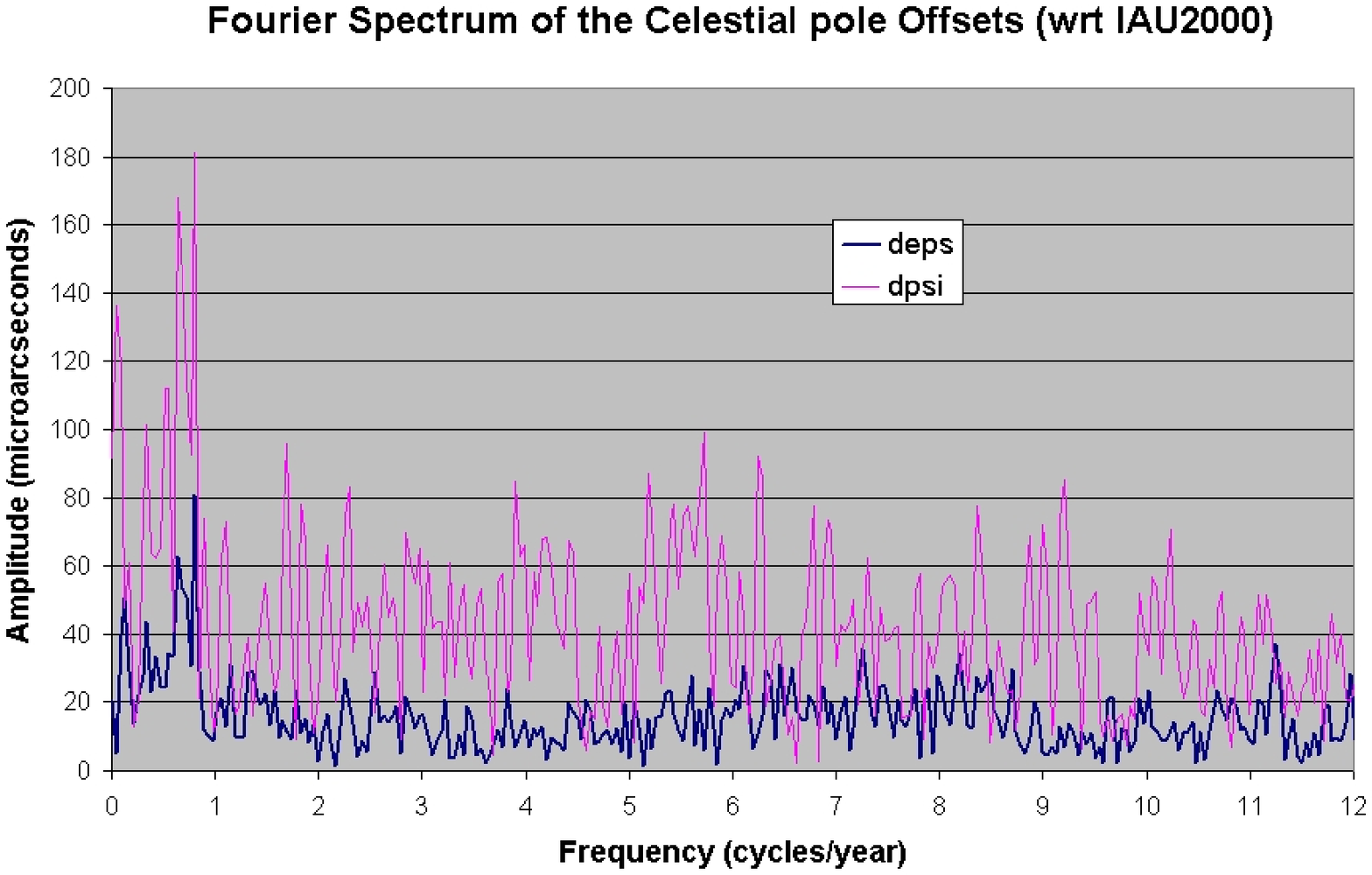}
\caption{Fourier spectrum of the celestial pole offsets obtained with OCCAM 6.1, with respect to IAU2000 model.}\label{fig:Fourier}
\end{center} 
\end{minipage}  
\end{figure}

In Figure \ref{fig:Xp}, we show the results obtained for the polar motion ($x_p$ component) compared to the IVS combined rapid solution (series ivs03r1e). This series was obtained using the IVS-R1 and R4 sessions, very good ones for determining the EOP, occuring after the year 2000. We can notice a shift in our series with respect to the IVS combined one. We wondered if this shift could be solved removing Tigoconception station. And we can realise that it is in better agreement after removing it (see Fig. \ref{fig:Xp}). This can be explained by the fact that we used ITRF2000 terrestrial reference frame, in which Tigoconception position is not well determined. But it still remains a difference at the end of the data time span. 
In Figure \ref{fig:Fourier}, we plotted the Fourier spectrum of the Celestial Pole Offsets (d$\psi$, d$\epsilon$) we obtained, with respect to the IAU2000 nutation model.  Depending on the data time span considered, we do not obtain the same periodical signals. The main ones are summarized in Table \ref{tab:FCN}, investigating the longest data time span possible (between 1982 and 2004). 
For further studies about the Free Core Nutation effect (FCN), we will investigate a wavelet analysis on the prograde and retrograde parts of the Earth nutations. In the future, we will be able to use also the intensive VLBI sessions in our computations.

\begin{table}[h]
\begin{center}
\begin{tabular}{l|rr}
Period    &    d$\epsilon$  &        d$\psi$  \\
\hline
450 days  &     81 $\mu$as  &    181 $\mu$as  \\
570 days  &     63 $\mu$as  &    168 $\mu$as  \\
\end{tabular}
\caption{Amplitudes of the biggest periodical signals for (d$\psi$,d$\epsilon$), with respect to IAU2000.}\label{tab:FCN}
\end{center}
\end{table}

\noindent {\large 4. REFERENCES}
{
\leftskip=5mm
\parindent=-5mm
\smallskip

Boehm, J., Schuh, H., 2004, "Vienna Mapping Functions in VLBI analyses", Geophys. Res. Lett., 31, doi: 10.1029/2003GL018984.

Heinkelmann, R., Boehm, J., Schuh, H., 2005, "IVS long-term series of tropospheric parameters", 6th IVS Analysis Workshop, April 21-23 2005, Noto Observatory, Sicily, Italy.

Koch, K.R., 1997, "Parameterschatzung und Hypothesentests in linearen Modellen", 3. Auflage, Ferdinand Dummlers Verlag, Bonn, pp. 381.

Mathews, P.M., Herring, T.A., Buffet, B.A., 2002, "Modeling of nutation and precession: New nutation series for nonrigid Earth and insights into the Earths's interior", \jgr (Solid Earth), 107, doi: 10.1029/2001JB000390.

Lieske, J.H., Lederle, T., Fricke, W., and Morando, B., 1977, "Expressions for the precession quantities based upon the IAU 1976 system of astronomical constants", \aa 58, pp. 1--16.

}

\end{document}